\documentclass{article}
\usepackage{psfrag,graphicx,a4,cite}
\begin{document}

\markboth{A.A. Lagutin, A.G. Tyumentsev, A.V. Yushkov}{On the deficit of calculated muon flux}

\title{On the deficit of calculated muon flux at sea level 
for energies $>100$ GeV} 

\author{A.A.~LAGUTIN$^\ast$,
A.G.~TYUMENTSEV$^\dagger$, A.V.~YUSHKOV$^\ddagger$} 
\date{Theoretical Physics Department, Altai State University,\\
61 Lenin Avenue, 656049 Barnaul, Russia\\ $^\ast$lagutin@theory.dcn-asu.ru\\ 
$^\dagger$tyumentsev@theory.dcn-asu.ru\\ $^\ddagger$yushkov@theory.dcn-asu.ru} 

\maketitle

\begin{abstract}
In this paper we discuss the problem, why the use of the direct data on primary 
nuclei spectra together with the modern hadronic interaction models leads to 
significant deficit of computed vertical muon flux at sea level for energies 
$>100$~GeV. We suggest, that to find out the source of this inconsistency it is 
necessary to perform an analysis of sensitivity of emulsion chamber data to 
variations of hadron-nucleus interaction characteristics. Such analysis will 
give more ground for discussion of adequacy of the up-to-date interaction 
models and of mutual compatibility of primary nuclei spectra, obtained in 
direct and EAS experiments. 
\end{abstract}


The accuracy, achieved in the last decade in measurements of primary cosmic ray 
(PCR) flux and in description of hadronic interactions, makes it possible to 
perform their consistency check. It seems that the most clear and easy way to 
do it lies via the calculation of uncorrelated muon flux, since it contains 
information both on the total PCR all-nucleon flux and high-energy interaction 
characteristics and it is rather precisely studied up to the energy of 10 TeV. 
The majority of such calculations, performed in the last 20 years, were made 
with the use of PCR spectra, inconsistent with the present data of direct 
measurements, and applied purely phenomenological, i.e. approximating 
accelerator data, nuclear interaction models (see Ref.~\cite{ya2004} for more 
details). These calculations did not find any difficulties in reproduction of 
the experimental data on the muon flux. In contrast to them, our results, 
presented in Ref.~\cite{ya2004}, clearly indicate, that the total vertical muon 
flux at sea level, resulting from the current balloon and satellite data on PCR 
spectra, is at the least 30--40\% deficient for the energies $>100$~GeV, 
regardless of interaction model applied: QGSJET\cite{qgsjet} or 
VENUS\cite{venus} (the latter one provides the largest number of muons at the 
sea level in comparison with other models, included in CORSIKA\cite{corsika}). 
The same conclusion can be drawn from the calculations in 
Refs.~\cite{lecoultre01,aires_caprice,chirkin_corsika}, also relying on the 
up-to-date information. The given discrepancy may be attributed only to the 
incorrectness of the interaction models, applied either for calculation of 
secondary particle fluxes in the atmosphere or for simulation of cascade 
processes in emulsion chambers (EC) in balloon experiments. The first group 
models (QGSJET, VENUS, DPMJET, SYBILL, NEXUS) are widely used in EAS 
experiments and the influence of differences between them on shower development 
is well understood. The detailed characteristics of interaction codes, applied 
in the direct measurements, except RUNJOB, are not a matter of common 
knowledge, and their consistency with just listed models was never 
investigated. Apparently, that these questions are of great importance, 
especially if to take into account, that due to high registration threshold the 
EC technique is sensitive to very scarcely known parameters of fragmentation 
particles. Possibly, that the mutual inconsistencies in these models may be the 
cause of 100\% scatter of the data on the PCR nuclei spectra with $Z\geq2$. To 
find out, whether it is so or not, it is necessary to perform an analysis of 
sensitivity of the EC data to the variations of hadronic cross-sections with 
the use of thoroughly developed modern interaction models, describing events on 
the basis of fundamental physical principles in the whole phase space, 
including very forward region, unattainable for study at the existing 
accelerators. Until such analysis is done, it is impossible to say, if the muon 
deficit should be attributed singly to the underestimation of primary particle 
energy in the direct experiments or it is needed to correct interaction models 
with an automatically following from this recalculation of the EC data. If to 
consider the first possibility, then, as it is discussed in our 
paper\cite{ya2004}, the lack of muons may be related in the large part only to 
the underestimation of flux of primary protons. 
\begin{figure}
\centering
\psfrag{Emu, GeV}[c][c][0.8]{$E_\mu$, GeV}
\psfrag{SmuE2}[c][c][0.8]{$S_\mu\times E_\mu^2$, GeV/($\mathrm{cm}^2\cdot\mathrm{s}\cdot\mathrm{sr}$)}
\psfrag{1e-03}[r][r][0.8]{$10^{-3}$}
\psfrag{1e-04}[r][r][0.8]{$10^{-4}$}
\psfrag{1e-02}[r][r][0.8]{$10^{-2}$}
\psfrag{10}[r][r][0.8]{$10$}
\psfrag{100}[r][r][0.8]{$100$}
\psfrag{m+}[r][r][0.8]{$\mu^+$}
\psfrag{m-}[r][r][0.8]{$\mu^-$}
\psfrag{BESS-TeV}[r][r][0.75]{BESS-TeV}
\psfrag{L3+C}[r][r][0.75]{L3+C}
\psfrag{CAPRICE94}[r][r][0.75]{CAPRICE}
\includegraphics[width=0.5\textwidth]{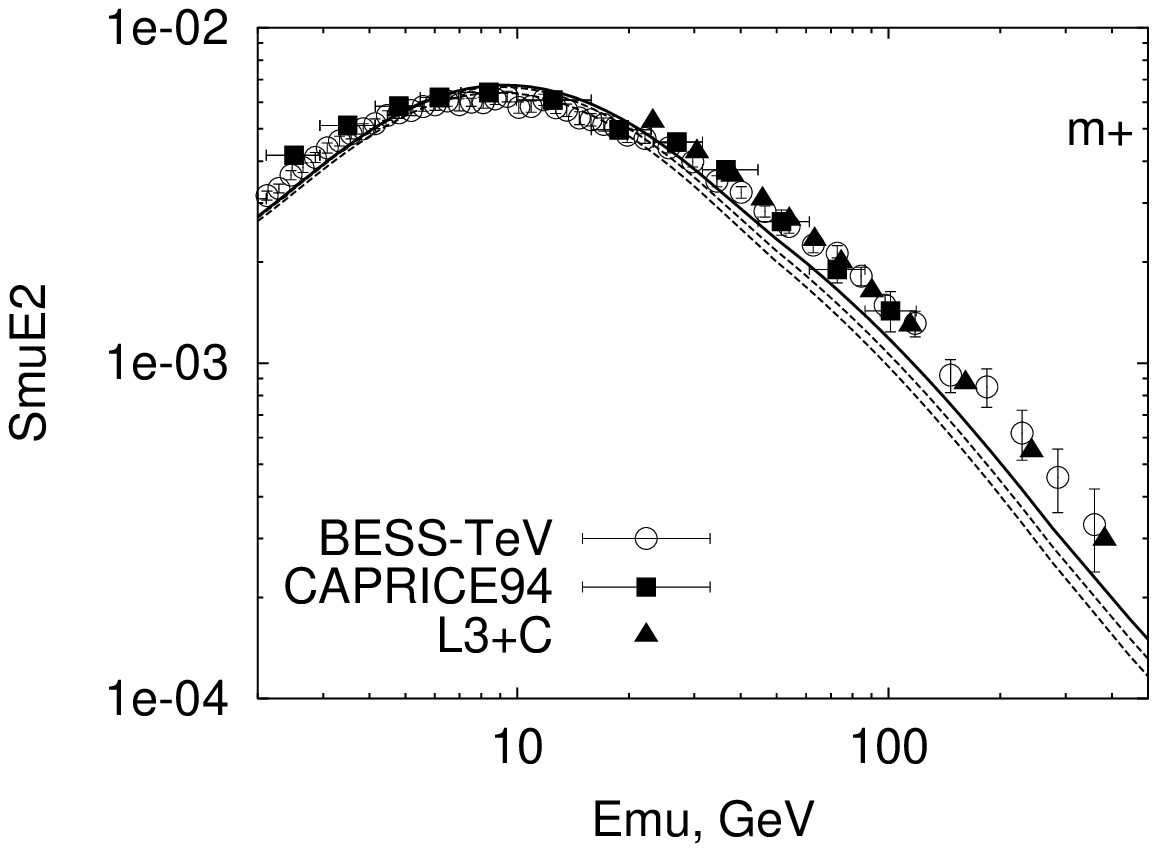}\hfill
\includegraphics[width=0.5\textwidth]{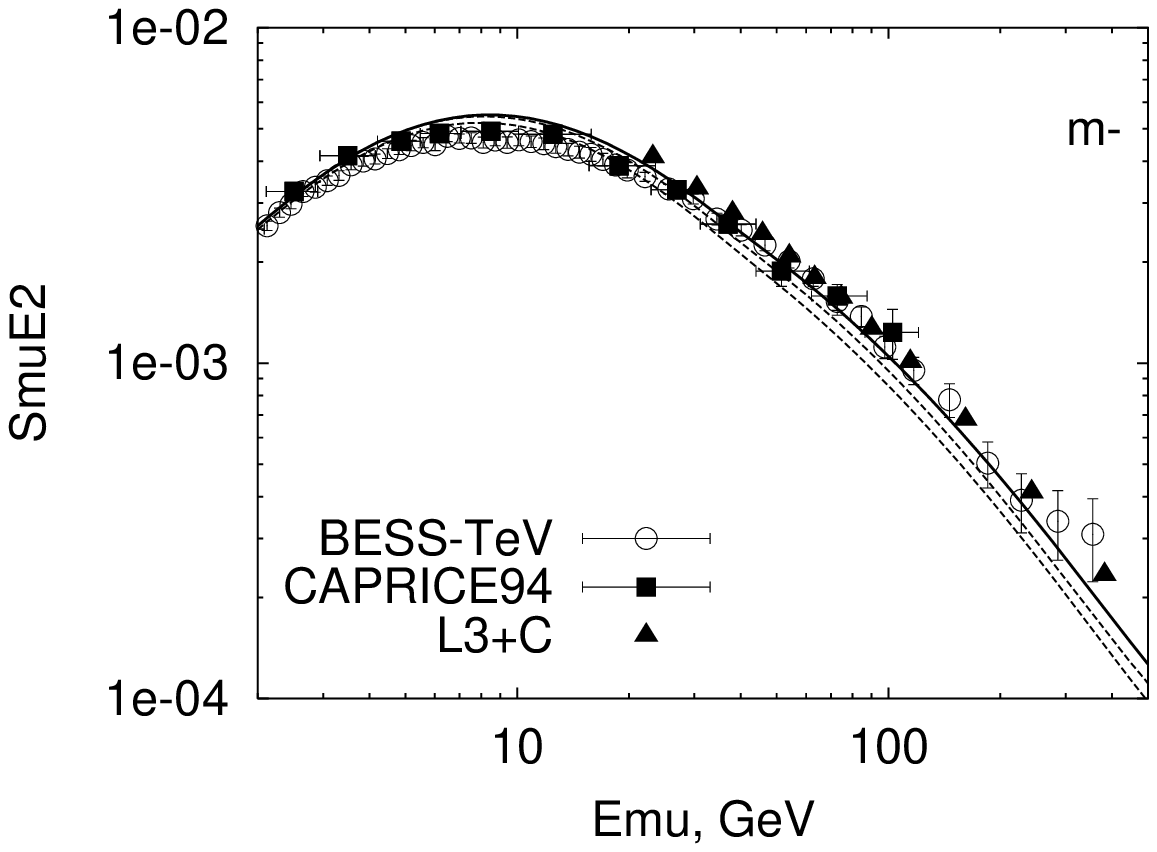}
\caption{Differential spectra of $\mu^+$ (left) and $\mu^-$ (right) at sea level.
This work calculation with CORSIKA/QGSJET (GHEISHA, $E_\mathrm{lab}<80$~GeV): 
solid line~---~for PCR fluxes in anomalous diffusion model${}^1$, dashed 
lines~---~for PCR spectra approximations, proposed in Ref.~9 for low and high 
helium flux fits. Experiments: CAPRICE~---~Ref.~10, BESS-TeV~---~Ref.~11, 
L3+C~---~Ref.~12.} 
\end{figure}
One may expect a confirmation of this conclusion from calculations of muon 
charge ratio, since it is sensitive to the chemical composition of PCR, but at 
the present state of the art such analysis is very speculative. For example, 
our computations of $\mu^+$ and $\mu^-$ fluxes, performed with the use of 
QGSJET (see Fig.~1), show, that the deficit of total muon flux for this 
interaction model is mostly due to a lack of positive muons. But it can not be 
interpreted as a proof of underestimation of primary proton flux, since its
enhancement would rise $\mu^+$ and $\mu^-$ intensities almost in equal amounts, 
thus leading to contradiction with the data on negative muons. In the given 
case $\mu^+$ deficit must be related to the fact, that the QGSJET model 
predicts low, in comparison with experiments and other interaction models, 
value of muon charge ratio\cite{unger_ecrs2004}. 

Consequently, we come to conclusion, that the systematic error in muon flux and 
charge ratio calculations is a sum of errors of two different interaction 
models, applied for simulation of cascades in EC and atmosphere, and it is 
impossible to separate them. Evidently, that to reduce error single model must 
be used for estimation of primary and secondary particle fluxes. For this 
purpose it is preferable to take not phenomenological and widely applied in EAS 
experiments interaction models. It is needed not only to get a consistent 
picture on the CR fluxes from the top of the atmosphere to the sea level, but 
also would allow to check the correctness of the underlying physics for these 
models in the high $x=E'/E$ region, which plays an important role both in 
interpretation of the EC data and in formation of muon spectrum, and to make 
more justified comparisons of direct and EAS PCR flux measurements (see such 
investigation in\cite{hoerandel03_knee,hoerandel03_cross}). Without this, the 
balloon data on PCR spectra can not be considered as the normalization 
standard, especially near the ``knee'', and it is impossible to understand, how 
the interaction models should be corrected. 

\section*{Acknowledgements}

This work is supported in part by the grants of the UR program No. 02.01.001 
and RFFI No. 04--02--16724.

\end{document}